\documentclass[aps,pre,twocolumn]{revtex4-1}
\pdfoutput=1
\usepackage{graphicx}
\usepackage{lipsum}

\usepackage{enumerate}

\usepackage{amssymb,amsfonts,amsmath,color}

\begin{document}

\title{Front propagation and  effect of memory in stochastic desertification models with an absorbing state}

\author{Dor Herman}
\affiliation{Department of Physics, Bar-Ilan University,
Ramat-Gan IL52900, Israel}

\author{Nadav M. Shnerb}
\affiliation{Department of Physics, Bar-Ilan University,
Ramat-Gan IL52900, Israel}


\begin{abstract}
\noindent
Desertification in dryland ecosystems is considered to be a major environmental threat that may lead to devastating consequences. The concern increases when the system admits two alternative steady states and the transition is abrupt and irreversible (catastrophic shift). However, recent studies show that the inherent stochasticity of the birth-death process, when superimposed on the presence of an absorbing state, may lead to a continuous (second order) transition even if the deterministic dynamics supports a catastrophic transition. Following these works we present here a numerical study of a one-dimensional stochastic desertification model, where the deterministic predictions are confronted with the observed dynamics. Our results suggest that a stochastic spatial system allows for  a propagating front only when its active phase invades the inactive (desert) one. In the extinction phase one observes  transient front propagation followed by a global collapse. In the presence of a seed bank the vegetation state  is shown to be more robust against demographic stochasticity, but the transition in that case still belongs to the directed percolation equivalence class.
\end{abstract}
\maketitle

\section{Introduction}

Systems governed by nonlinear dynamics may support two or many fixed points for the same set of parameters. Slow variation of the external parameters may lead to an abrupt change of the state of the system (catastrophic shift) at the tipping point, where one of the equilibrium states loses its stability. In ecological systems, these shifts are often harmful and  may cause a loss of bioproductivity and biodiversity, which, in turn, negatively affect ecosystem function and  stability~\cite{petraitis2013multiple,scheffer2001catastrophic}. Therefore, the possibility that ecosystems may undergo such an irreversible transition  in response to small and slow environmental variations raises much concern \cite{staal2015synergistic, nes2014tipping,reyer2015forest,baudena2015forests,martinez2016drought,eby2017alternative}.

\begin{figure}
\centerline{\includegraphics[width=8cm]{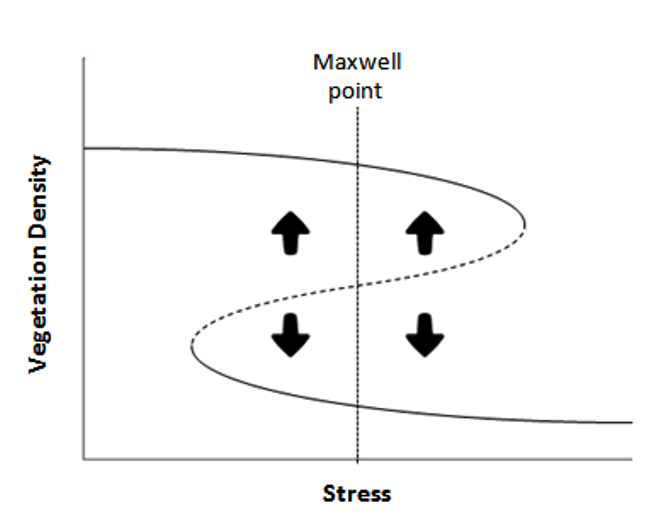}}
\vspace{-0.cm}
\caption{The different regime-shift scenarios for a vegetation system with strong positive feedback are illustrated in this cartoon, where full lines represent stable fixed points of the vegetation dynamics and the dashed line the unstable fixed point between them.  For very low stress the system admits only one stable state with high  vegetation density, while for very high stress there is a region with only one, low density state. For intermediate stress  the system supports, for each stress value, two alternative states separated by an unstable fixed point. Below the unstable fixed point the vegetation density is too small and positive feedback is too weak therefor the density shrinks, while above this point the density increases (bold arrows). In well-mixed situations the system stays in one of its stable states until it hits the corresponding tipping point where it undergoes a catastrophic shift (saddle-node bifurcation). In spatial domains the low-vegetation (desert) phase invades the high-vegetation one to the right of the Maxwell (stall) point, while the high vegetation invades the desert if the stress is smaller than its Maxwell point value~\cite{bel2012gradual}.  }
 \label{fig1}
\end{figure}

Catastrophic shifts are considered as an important factor in many studies of  transitions between various vegetation regimes, including the destructive process of desertification~\cite{foley2003regime,janssen2008microscale,sankaran2005determinants}.
Many works suggest a positive correlation between the local vegetation density and its growth rate as a result of various positive feedback mechanisms like shading, root augmentation, infiltration rates, and fire cycles~\cite{meron2012pattern,gilad2004ecosystem,adams2013mega,tredennick2015effects}. If the positive feedback is strong enough the system may support alternative steady states. The transition from one state to another may take place abruptly beyond the tipping point, or (in spatial domains) gradually, with the stable phase invading the metastable one \cite{durrett1994importance,bel2012gradual}. A general cartoon that illustrates the  possible scenarios is shown in Figure \ref{fig1}.

However, ecological systems in general are known to be very noisy, affected by all kinds of stochastic processes~\cite{lande2003stochastic,kalyuzhny2014niche,kalyuzhny2014temporal}. In particular, the stochasticity involved in the birth-death process of individuals (demographic noise) may modify the characteristics of the transition, as it supports an absorbing state. Unlike other types of stochasticity that only lead to a shift in the location of the Maxwell point \cite{khain2011fluctuations}, the transition in the presence of an absorbing state  may be continuous (second order) and reversible, usually belong to the directed percolation equivalence class \cite{kockelkoren2003absorbing,bonachela2012patchiness,agranovich2012predator,fried2013damage}. Following these observations, a minimal model for desertification transition in one~\cite{weissmann2014stochastic} and two~\cite{martin2015eluding} dimensions have been considered in recent years. It turns out that in $1d$ the transition is always continuous, while in two spatial dimensions the characteristics of the transition depends on the strength of the noise. For weak noise one observes a catastrophic shift, while under strong noise the transition is again continuous. In both cases, the continuous (second order) transition  belongs to the directed-percolation equivalence class.

In this paper we would like to extend the work of~\cite{weissmann2014stochastic}. We consider the one-dimensional case, and focus our attention on two aspect of the relationships between the deterministic and the stochastic dynamics: the propagation of fronts when one phase invades the other, and the effect of memory on the transition between the active and the extinction phase.

In the next section we study the remnants of the deterministic behavior in the stochastic system. Deterministically, the Maxwell (stall) point (see Figure \ref{fig1})  separates the region where desert invades vegetation from the region where vegetation invades the desert. Under demographic stochasticity, the transition point is inevitably to the left of the Maxwell point~\cite{weissmann2014stochastic}, and above the transition  the vegetation eventually collapses (in regions II, III and IV indicated in Fig. \ref{fig2}). Still, we would like to check the possibility of a  transient invasion (of desert into vegetation or vegetation into desert - see below) in these regimes of parameters.

 In section 3 we address a more practical question, the effect of long term memory, in the context of a model that includes a seed bank in the soil. We will show that memory makes the system more resilient against the effects of demographic stochasticity so the shift of the transition point is larger than the shift predicted by the deterministic model but the transition is still continuous and directed-percolation like.  This section is followed by a general discussion and conclusions.

\section{Front propagation and invasion in  the stochastic model}

The most popular model for a catastrophic desertification is a version of the Ginzburg-Landau (GL) diffusion-reaction model,
\begin{equation} \label{equation:DesertificationNoSeeds}
\centering
\dot{\varphi}\left(x,t\right) = D\nabla^{2}\varphi\left(x,t\right) - \alpha\varphi\left(x,t\right) + \beta\varphi^{2}\left(x,t\right) - \gamma \varphi^{3}\left(x,t\right)
\end{equation}
Here $\varphi$ represents the vegetation density,  $\mathnormal{D}$ is the diffusion coefficient, $\alpha$ is the stress parameter (measuring  the harshness of the environmental conditions), $\beta$ sets the scale of  the positive feedback and $\gamma$ represents the finite carrying capacity of the system.

\begin{figure}
 	\centerline{\includegraphics[width=8cm]{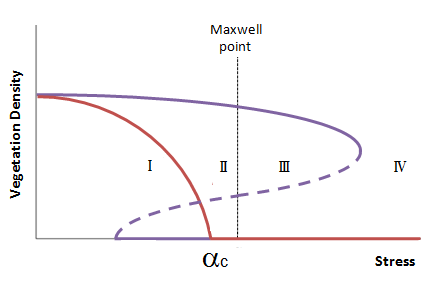}}
 	\vspace{-0.cm}
 	\caption{An illustration of the relationships between the stochastic (red) and the deterministic (purple) states of the system as  obtained in \cite{weissmann2014stochastic}. Unlike the deterministic dynamics illustrated in Fig. \ref{fig1}, under demographic stochasticity the transition is continuous (second order) and must occur to the left of the Maxwell point. In region II the deterministic theory predicts that the active (vegetation) phase invades the desert, while in region III the desert invades  vegetation. }
 	\label{fig2}
 \end{figure}

 As the stress parameter $\alpha$ is increased the environmental conditions deteriorate and the vegetation density reduces. The first, transcritical, bifurcation (the left tipping point in Figure \ref{fig2}) occurs at $\alpha =0$, where the state $\varphi =0$ becomes stable. The right tipping point (a saddle-node bifurcation) occurs at $\alpha = \beta^{2}/4\gamma$, and if the system reaches this point the vegetation collapses at once. The Maxwell (stall) point occurs at  $\alpha_{MP} = 2\beta^{2}/9\gamma$: to the left of this point vegetation invades the desert, while to its right desert invades vegetation. The velocity of the invasive front is zero at the Maxwell point and is linearly proportional to  $\alpha_{MP}-\alpha$ in the vicinity of the Maxwell point.

 Figure \ref{fig2} illustrates the relationships between the stochastic and the deterministic transition dynamics as obtained in \cite{weissmann2014stochastic}. In the presence of demographic noise and an absorbing state the transition is always continuous, occurs  to the left of the Maxwell point and belongs to the directed percolation equivalence class. Accordingly, between the stochastic extinction point $\alpha_c$ and the Maxwell point there is a region of parameters where the deterministic theory predicts invasion of vegetation while under stochastic dynamics the vegetation goes extinct (region II in Fig. \ref{fig2}). For $\alpha > \alpha_{MP}$ one finds another region, where under deterministic dynamics the desert invades (region III). Our aim here is to check for the traces of this deterministic behaviors when the system is  stochastic.

 Even in the quasi-deterministic limit, where the effect of the absorbing state is negligible, demographic stochasticity affects the width and the velocity of the invasion front \cite{meerson2011velocity,khain2013velocity}. As was shown in \cite{martin2015eluding}, under demographic noise the renormalized value of $\beta$ (the positive feedback term) decreases as the length scale increases, meaning that at the deterministic Maxwell (stall) point, desert still invades. Here we present a few simulations done for strong and intermediate levels of noise, where the effect of the absorbing state is pronounced.

\begin{figure}
 	\centerline{\includegraphics[width=8cm]{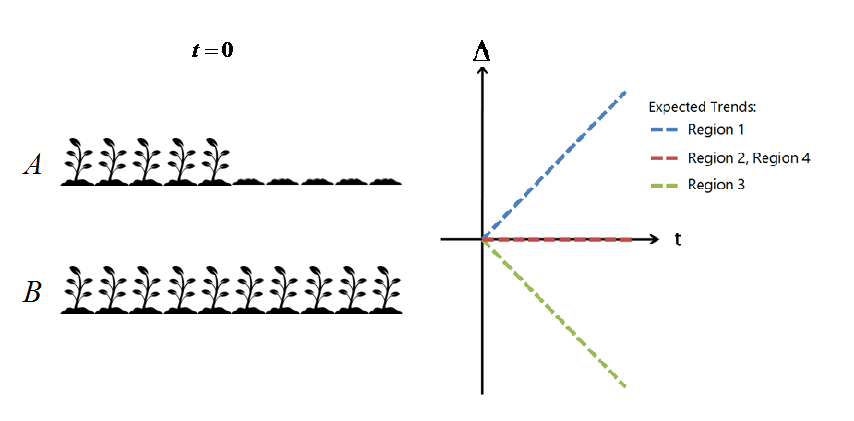}}
 	\vspace{-0.cm}
 	\caption{Identification of  spatial effects by comparison of  the time evolution of two different types of initial conditions. We have traced the overall vegetation in a one dimensional system that was prepared  in a homogenous spatial state that corresponds to the stable vegetation fixed point of the deterministic dynamics (B). This value was compared with  another system (A), in which this state covers half of the system while the other half is in  bare soil (desert). $\Delta$, as explained in the text, is the difference between the total biomass of $A$ and half of the biomass of $B$. If the vegetation state invades the desert, even as a transient, one expects a growth of $\Delta$ in time, as illustrated by the blue dashed line in the right panel. The opposite scenario, an invasion of the desert, will yield a curve with negative slope (green). If spatial invasion plays no essential  role, $\Delta$ is expected to be time independent (red). }
 	\label{fig3}
 \end{figure}

 In the simulation we have compared the outcomes of the stochastic dynamics with different initial conditions (see Fig. \ref{fig3}). In case A the initial condition is inhomogeneous, half the system ($0 \le x \le L$) is in the vegetation state (as predicted by the deterministic theory)  and the other half ($L \le x \le 2L$) in the absorbing state. In the homogeneous case B, the system is initially in the vegetation state for all $x$. We then tracked the value of
  \begin{equation}
  \Delta(t) \equiv   \int \varphi_A(x,t)\  dx -\frac{1}{2}  \int \varphi_B(x,t)\  dx 
  \end{equation}
  through time.  In the  region of parameters  that corresponds to phase II  one may suppose that for some time the deterministic prediction still holds and vegetation invades the desert, so $\Delta$ will increase in time. This increase has to be a transient since eventually the system must collapse to the absorbing state, but it may manifest itself for a while. Similarly, if the desert invades deterministically  (phase III) $\Delta$ should decrease in time. Finally if spatial invasion have no significant role, $\Delta$ would be time independent. These scenarios are illustrated in Fig. \ref{fig3}.

 \begin{figure}
 	\centerline{\includegraphics[width=8cm]{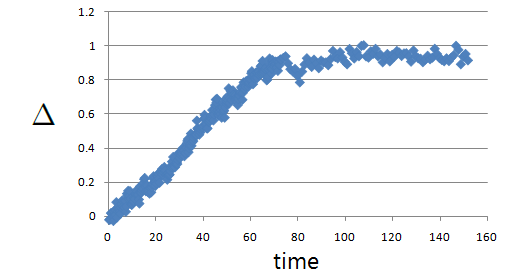}}
 	\vspace{-0.cm}
 	\caption{$\Delta$ as a function of $t$ in the active case (phase I, $\alpha = -1.8$ where the stochastic transition is around $\alpha_c \sim -0.7$). As expected $\Delta$ grow and reaches equilibrium at long times. Biomass density is measured in units of the deterministic fixed point, so the value it approaches is slightly smaller than one. The line presented here is an average over the time evolution of $100$ systems.}
 	\label{fig4}
 \end{figure}

\begin{figure}
 	\centerline{\includegraphics[width=8cm]{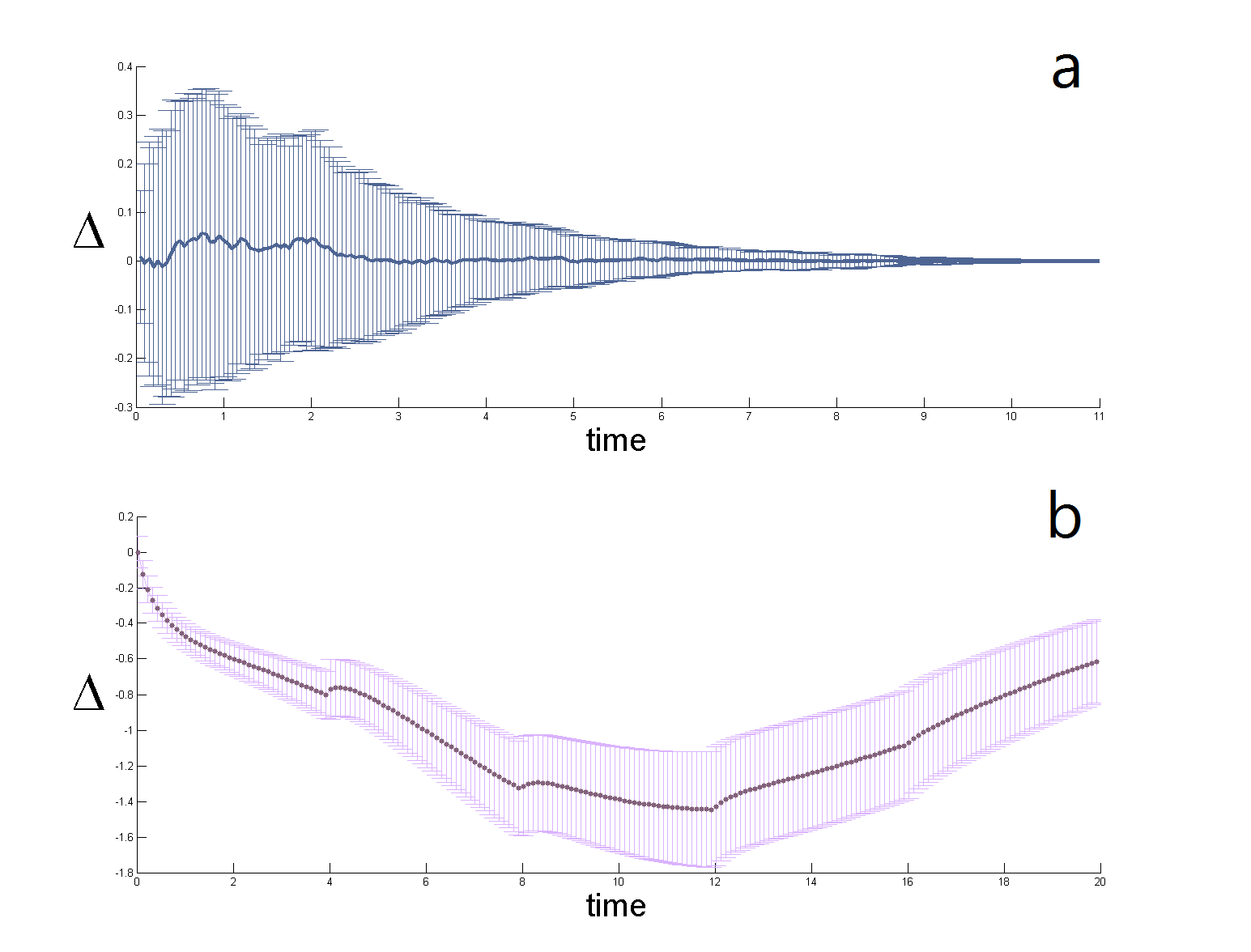}}
 	\vspace{-0.cm}
 	\caption{The average value of $\Delta$ (dark circles) and its (one standard deviation) error bars  are plotted as a function of $t$ where $\alpha = 1.97$ is in phase III. For small noise  ($\zeta = 30$, panel a) one cannot detect deviations between $\Delta$ and zero through the (one standard deviation) error bars. For weak noise (Panel b, $\zeta = 10^6$) the decay and growth are pronounced. Similar results were obtained using a different simulation technique for the same system, see appendix A.}
 	\label{fig5}
 \end{figure}

\begin{figure}[h]
 	\centerline{\includegraphics[width=8cm]{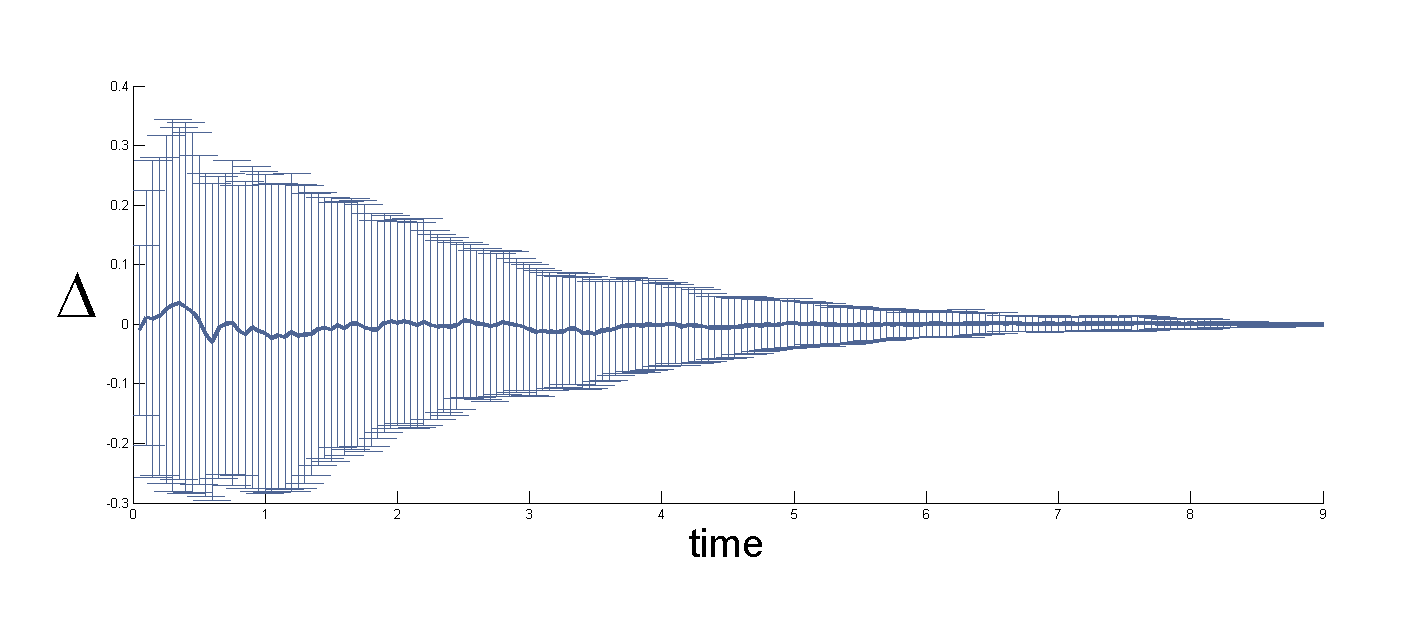}}
 	\vspace{-0.cm}
 	\caption{$\Delta$ as a function of $t$ when $\alpha=2.1$ is  in region IV, where the vegetation collapses and the time to local extinction is logarithmic in the population size. The lines are  an average over the time evolution of $100$ systems, and the error bars represent one standard deviation.}
 	\label{fig7}
 \end{figure}

 The vegetation system was simulated using the algorithm presented in \cite{weissmann2014stochastic}. The deterministic equation (\ref{equation:DesertificationNoSeeds}), with $D=0.2$, $\beta = 0.4$, $\gamma = 0.02$, and different values of  $\alpha$, was  integrated (using the Euler method, timesteps of $0.001$) for a one dimensional lattice with $2L=750$ and reflecting (no-flux) boundary conditions. Demographic noise was introduced into the dynamics by replacing, every $\zeta$ units of time,  the value of $\varphi(x,t)$ by an integer $m(x,t)$, drawn from a Poisson distribution,
 \begin{equation}
 P(m) = e^{-\varphi(x)} \frac{\varphi^m(x)}{m!}.
 \end{equation}
 We have used asynchronous update of the lattice. For the simulation in phase I, II and IV,   $\zeta =30$ was taken, while in region III we implemented different levels of noise as will be explained below. The results are shown in Figures \ref{fig4}, \ref{fig5},  and \ref{fig7}.

 In the active phase of the stochastic dynamics (region I of Fig. \ref{fig2}) the vegetation invades the desert as expected (Fig. \ref{fig4}), and $\Delta$ grows in time until it reaches an equilibrium value. This equilibrium value is smaller than the prediction of the deterministic theory, as one expects for a stochastic system with an absorbing state.
 The invasion velocity should decrease as $\alpha$ increases and it reaches zero at the transition point. As demonstrated in \cite{weissmann2014stochastic}, the transition occurs to the left of the Maxwell point and belongs to the directed percolation equivalence class, so the front velocity should vanish at this transition point (for details, see \cite{kessler2012scaling}), leaving a region (II) where the deterministic front velocity is finite (vegetation invades desert) but under stochasticity the  vegetation cannot invade.

 Before we consider region II, let us focus our attention on the third region, where deterministically the desert invades. Since this region is in the stochastic extinction phase, the vegetation will collapse in the long run even if the initial conditions are homogenous and active (i.e., correspond to case B of figure \ref{fig3}). This collapse will take place in two steps: first, the stochastic fluctuations must generate a spatial region where the vegetation density is low enough (i.e., it reaches the basin of attraction of the absorbing state) and then this desert "nucleus" invades the rest of the system with the (renormalized \cite{khain2013velocity}) Ginzburg-Landau velocity. Before this nucleation, one can see the desert invade the vegetation, while above this typical nucleation time the vegetation collapses everywhere and $\Delta$ approaches zero.

  This behavior is seen in  Figure \ref{fig5}. If  the noise is relatively strong and the nucleation time is short,  one observes no effect of invasion. For the same set of parameters but with weaker noise (larger $\zeta$),  $\Delta$ first decreases  (indicating invasion by the desert) and then grows back towards zero. The same behavior is seen in Fig. \ref{fig_MP} below, where a different algorithm was used (see Appendix A).

Phase III admits a well defined semi-classical limit. While the nucleation time increases exponentially with $\varphi(x)$, the inverse of the noise level at a site, the corrections to the front velocity scale like $1/\varphi$ \cite{khain2013velocity} so as $\varphi$ increases one can observe the deterministic behavior for exponentially long times.
Although the system is formally in the stochastic extinction phase, in every finite sample the effect of deterministic invasion will be much stronger than the accumulation of stochastic local extinctions.

At the Maxwell point the deterministic front velocity is zero, but under demographic noise the desert still invades vegetation since (as explained above) the renormalized positive feedback term is smaller than its bare value. This behavior was demonstrated in \cite{khain2013velocity}. We failed to simulate our system at the stall point since the noise was too large, but in Appendix A we present results that were obtained using a different numerical technique and support this conclusion.

This brings us to back to phase II, where deterministically the vegetation invades but stochastically the system is in the extinction phase. This region has no semiclassical limit: as $\varphi$ (or $\zeta$) increases (the demographic noise decreases) the extinction point approaches the Maxwell point and this phase disappears. Since the desert invades at the Maxwell point it must invade with finite velocity to the left of this point, but now the invasion is only due to stochastic effects (without stochasticity the active state invades) and for any fixed value of $\alpha$ the velocity changes  sign as $\varphi$ increases. To put it another way, for any fixed strength of demographic stochasticity there is, somewhere in region II,  a critical value, $\alpha_{SMP}$, the stochastic Maxwell point, and to the left of this point the desert does not invade.

The only question that remains is about the relationship between $\alpha_c$, the extinction transition point (above it vegetation no longer invades the desert) and the stochastic Maxwell point $\alpha_{SMP}$ (below it desert cannot invade vegetation). Do these two point coincide? Unfortunately, we did not succeeded in answering this question using our numerics since fluctuations in the relevant regime are too strong.

Finally, in region IV (Figure \ref{fig7}) the decay of the active phase is deterministic so both systems show the same behavior, as expected.

%
%

\section{Memory effect and seed bank}

In standard vegetation systems the reproduction of individuals is aided by the dispersal of seeds. Seeds do not necessarily germinate when the environmental conditions allow it (e.g., during the first wet season); instead, they may enter a dormant state and germinate only after a few years. Seeds dormancy is a known bet-hedging strategy which is crucial to the ability of a species to expand and survive in harsh environmental conditions \cite{philippi1993bet,bewley1997seed}. From the modeler's perspective, this behavior introduces an  effective memory to the dynamics, as the "absorbing state" is not the state without vegetation but the state with no vegetation and  no seeds. In this section we present a numerical study of a model that takes into account this memory effect.

To do that, we have used one of the standard deterministic models of vegetation-seed dynamics (this specific version was adapted from Godoy, et al.  \cite{godoy2014phylogenetic}). Here, as before, $\varphi$ is the vegetation density that admits the same Ginzburg-Landau dynamics, and $\theta(x,t)$ is the density of seeds in the soil. Seeds germinate at a rate $g$  and die at a rate $\mu$. Finally,  $F$ is the rate (per biomass) in which the vegetation produces seeds.
\begin{widetext}
\begin{eqnarray} \label{equation:desertificationwithdeeds}
\dot{\varphi}\left(x,t\right) &=&  D\nabla^{2}\varphi\left(x,t\right) - \alpha\varphi\left(x,t\right) + \beta\varphi\left(x,t\right)^{2} - \gamma\varphi\left(x,t\right)^{3} + g\theta\left(x,t\right)
\\ \nonumber \dot{\theta}\left(x,t\right) &=& F\varphi\left(x,t\right) - \left(g + \mu\right)\theta\left(x,t\right)
\end{eqnarray}
\end{widetext}

\begin{figure}
 	\centerline{\includegraphics[width=8cm]{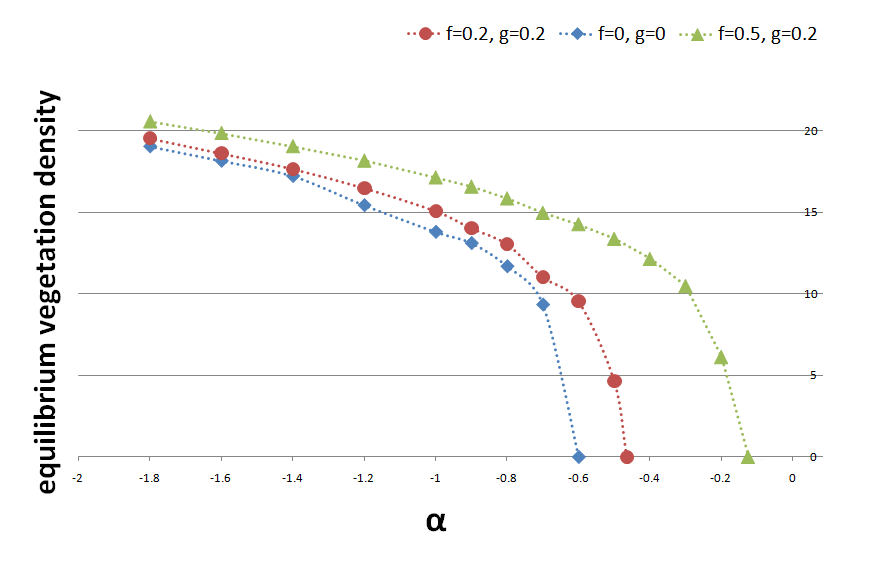}}
 	\vspace{-0.cm}
 	\caption{Equilibrium vegetation density as a function of $\alpha$ for the model (\ref{equation:desertificationwithdeeds}). In all systems $\beta = 0.4$, $\gamma = 0.02$, $D=0.2$ and $\zeta = 30$.  Results are presented for the case with no seeds $F=g=\mu=0$ (blue diamonds), for $F=g=0.2$ and $\mu=0.15$ (deterministic shift $gF/(g+\mu) = 0.11$, red dots) and for $F=0.5$, $g=0.2$ and $\mu = 0.15$ (deterministic shift $gF/(g+\mu) = 0.29$,  green triangles). The shift of the stochastic critical point is larger than the deterministic shift ($0.14$ instead of $0.11$ in the first case, $0.47$ instead of $0.29$ in the second case) due to the buffering effects of the seed dynamics.}
 	\label{fig8}
 \end{figure}

The deterministic steady states of (\ref{equation:desertificationwithdeeds}) are the sets of $\varphi,\theta$ where  both $\dot{\varphi}$ and $\dot{\theta}$ are zero. One may use that to solve for $\theta$ at the steady state, and by introducing this value of $\theta$ into the $\varphi$ equation it turns out that the deterministic phase diagram is shifted rigidly to the right by $gF/(g+\mu)$. However, as seen in Figure \ref{fig8}, the shift of the stochastic transition point is \emph{larger} than its deterministically expected value. This happens since the long-term memory of the system with seeds acts to buffer  the system against the effect of stochasticity.

\begin{figure}
 	\centerline{\includegraphics[width=8cm]{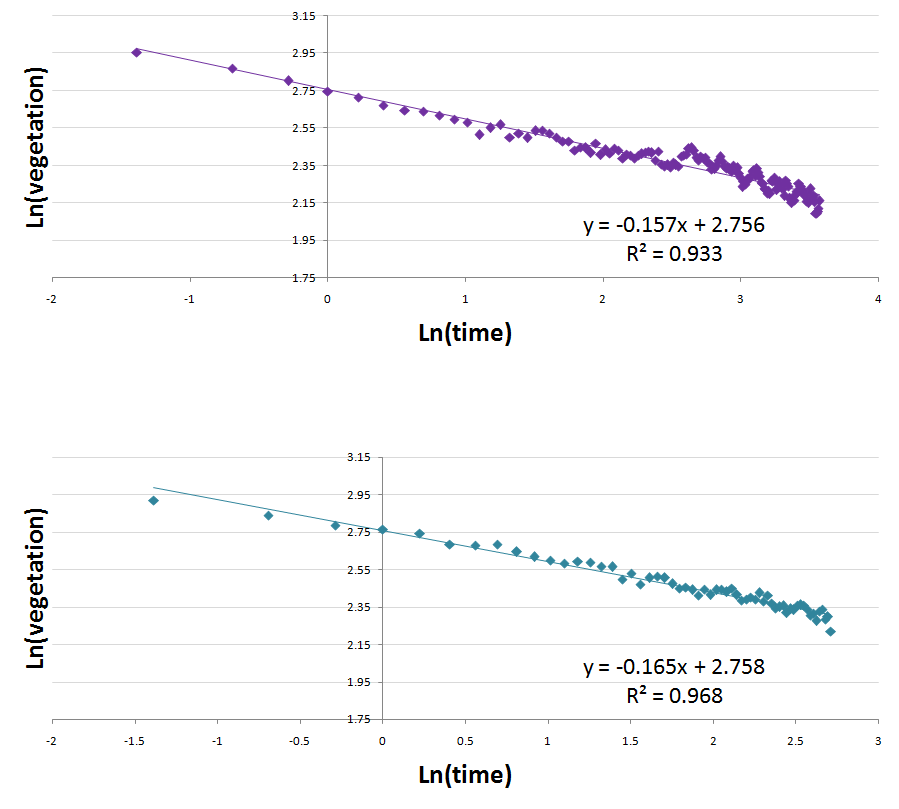}}
 	\vspace{-0.cm}
 	\caption{Vegetation density vs. time at the transition point, plotted on a double logarithmic scale. The decay fits quite nicely a straight line with slopes that are  very close to the values predicted for the directed percolation equivalence class, $\delta \sim 0.159$. The upper panel was obtained from the model with $F=0.2$, $g=0.2$ and $\mu = 0.15$, the lower panel for the model with $F=0.5$, $g=0.2$ and $\mu = 0.15$. The density vs. $\alpha$ diagram for these models is given in Fig. \ref{fig8}.   }
 	\label{fig9}
 \end{figure}

While the effect of memory increases the region of parameters that correspond to the active state, the characteristics of the transition itself appear to be the same. As explained in \cite{weissmann2014stochastic,martin2015eluding}, when the transition in the presence of  demographic noise falls into the directed percolation equivalence class the density of vegetation decays at the transition point like $\varphi \sim t^{-\delta}$, where in $1d$, $\delta \sim 0.159$. This behavior characterizes also the transition of the model with long-term memory and seeds, as indicated in Figure \ref{fig9}.

\section{Discussion}

In their very influential work, "The importance of being discrete (and spatial)" \cite{durrett1994importance}, Durrett and Levin considered two aspects of reality that may alter the predictions of a deterministic mean-field (well-mixed) theory in ecosystems. One is the invasion of a metastable by a stable phase, which yields a Ginzburg-Landau propagating front and eventually implies a gradual transition~\cite{bel2012gradual}, and the other is the effect of demographic stochasticity (discreteness of individuals) that may lead to sustainable coexistence even in the absence of an attractive fixed point~\cite{seri2012sustainability}.

Here we have shown that these two aspects of the dynamics - being spatial and being discrete - may also interfere with each other. When demographic stochasticity is superimposed on spatial dynamics, the net result is not Ginzburg-Landau invasion and  not the coexistance of the two states together. Instead the vegetation invasion dynamics is completely  identical with that of a Fisher front, i.e., of the invasion of a stable to an unstable (rather than metastable) state,  despite the fact that the deterministic system supports two attractive fixed points.

Seed dormancy is a known phenomenon in desert annuals \cite{philippi1993bet,bewley1997seed}, but in most cases it is considered as a bet-hedging strategy against \emph{environmental} variations. Here we have shown that such a strategy makes the system  also more resilient against the effect of demographic stochasticity, as it diminishes the influence of the absorbing state. On the other hand this memory effect via the seed bank do not change the properties of the transition itself. This may be related to the robustness of the directed percolation transition against spatio-temporal noise \cite{hinrichsen2000non}. The case of non-Markovian memory, like seeds with  power-law statistics for germination time, may yield, perhaps, a different type of transition, but the relevance of such a case to realistic systems is not  yet clear.

\appendix

\section{Front propagation at the Maxwell point}

In this appendix we describe a different numerical technique aimed at simulating the stochastic dynamics, and we implement this technique at  the deterministic Maxwell point. The numerical procedure used here is more realistic and more stable than the one used along this paper. However it does not allow one to vary the strength of the demographic stochasticity when all other parameters are kept fixed, so we use it only for this specific task.

For each site with $n$ individuals, the per-individual death rate is taken to be $$\mu_1 =\alpha+\gamma (n-1)(n-2)/6,$$ the per-individual birth rate is $$b_1 =\beta (n-1)/2,$$ and the per individual hopping rate is $D$. The overall "activity rate" per individual is thus, $$ r = \mu_1+b+1+D.$$

\begin{figure}
 	\centerline{\includegraphics[width=8cm]{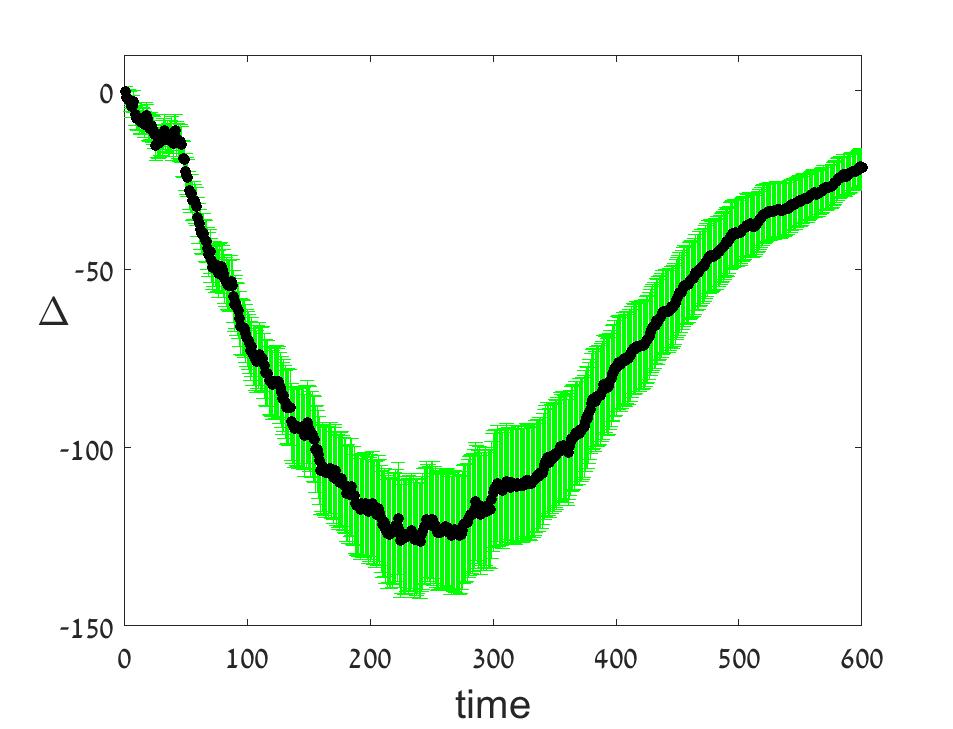}}
 	\vspace{-0.cm}
 	\caption{$\Delta$ vs. time as simulated by the multinomial algorithm for $\gamma = 0.001$, $\beta = 0.02$, $D=0.2$ and $\alpha = 0.136667$ (at the deterministic Maxwell point). A one dimensional system with $2L=300$ was simulated, using $\Delta t =0.1$. The deterministic solution for the steady state at this point is $n=42$. }
 	\label{fig_MP}
 \end{figure}

Accordingly, if $r \Delta t \ll 1 $ (where $\Delta t$ is the simulation time-step)  the chance that a single individuals does nothing during this period of time is $$Q = exp(-r \Delta t),$$ the chance that it give birth is $b_2 = (1-Q) b_1/r)$, death occurs with probability $d_2 = (1-Q) d_1/r$ and the probability of hopping is $h_2 = (1-Q)D/r$. By picking five numbers from a multinomial distribution for $n$ individuals with probabilities $[Q,h_2/2,h_2/2,b_2,d_2]$ we determined the number of individuals that stay inactive, jump to the left/right, give birth or die. The number $n$ is then updated to $n+2b_2$, and the hopping individuals are deposited in a different array and are added to $n$ only at the end of each turn (asynchronous updating) to avoid artificial drifts.

Figure \ref{fig_MP} shows $\Delta$ (obtained, as in the main part of this paper, by comparison between  two initial conditions, one homogenous and one with half vegetation) as a function of time at the Maxwell point, and one can see clearly that the desert invades at short times, then the vegetation in both systems collapse.

\begin{acknowledgments}
We acknowledge the support of the Israel
Science Foundation, grant no. $1427/15$. We thank Ehud Meron, David Kessler and Baruch Meerson for fruitful discussions and comments.
\end{acknowledgments}

\bibliography{dor_ref}
\end{document}